\documentclass[conference]{IEEEtran}

\usepackage{cite}
\usepackage{citesort}
\usepackage{graphicx}
\usepackage{amsmath}
\usepackage{amsfonts}
\usepackage{epstopdf}
\usepackage{xcolor}
\usepackage{amssymb,bm,upgreek}
\usepackage{algorithm}
\usepackage{algorithmic}
\usepackage{multirow}
\usepackage{transparent}

\usepackage{import}

\DeclareMathOperator\erf{erf}

\let\ss= \scriptscriptstyle
\newcommand{\RX}{\textnormal{RX}}
\newcommand{\R} {\textnormal{RX}}
\newcommand{\TX}{\textnormal{TX}}
\newcommand{\T}{\textnormal{TX}}
\newcommand{\FC} {\textnormal{FC}}
\newcommand{\trans}{\textrm{trans}}
\newcommand{\report}{\textrm{report}}
\newcommand{\metre}{\textnormal{m}}
\newcommand{\s}{\textnormal{s}}

\newcommand{\m}{\textnormal{m}}

\newcommand{\ob}{\textnormal{ob}}
\newcommand{\md}{\textnormal{md}}
\newcommand{\fa}{\textnormal{fa}}

\newcommand\ceil[1]{\lceil#1\rceil}

\begin{document}

\title{Distributed Cooperative Detection for Multi-Receiver Molecular Communication}

\author{\IEEEauthorblockN{Yuting Fang${}^\dag$, Adam Noel${}^\ddag$, Nan Yang${}^\dag$, Andrew W. Eckford${}^\sharp$, and Rodney A. Kennedy${}^\dag$}
\IEEEauthorblockA{${}^\dag$Research School of Engineering, Australian National University, Canberra, ACT, Australia\\}
\IEEEauthorblockA{${}^\ddag$Department of Computer Science and Operations Research, Universit\'{e} de Montr\'{e}al, Montr\'{e}al, Qu\'{e}bec, Canada\\}
\IEEEauthorblockA{${}^\sharp$Department of Electrical Engineering and Computer Science, York University, Toronto, Ontario, Canada}}


\markboth{Submitted to IEEE GLOBECOM 2016}{Yuting \MakeLowercase{\textit{et al.}}: Distributed Cooperative Detection for Multi-Receiver Molecular Communication}

\maketitle

\begin{abstract}
In this paper, a cooperative diffusion-based molecular communication system is considered where distributed receivers collaboratively determine a transmitter's signal. In this system, the receivers first make local hard decisions about the current transmitted bit and then report these decisions to a fusion center (FC). The FC combines the local hard decisions to make a global decision using an $N$-out-of-$K$ fusion rule. Asymmetric and symmetric topologies are considered and for each topology, two reporting scenarios, namely, perfect reporting and noisy reporting, are addressed. Closed-form analytical expressions for the expected global error probability are derived for all considered topologies and scenarios. Numerical and simulation results show that system reliability can be greatly improved by combining the detection information of distributed receivers.
\end{abstract}

\IEEEpeerreviewmaketitle

\section{Introduction}\label{sec:intro}

Over the past decades there have been considerable advancements in designing and engineering nanoscale ($<0.1{\mu}\metre$) and microscale ($0.1$ to $100 {\mu}\metre$) devices. These devices can be interconnected to execute complex tasks, e.g., intra-body drug delivery, in a cooperative manner. The resulting network, i.e., nanonetwork, is envisaged to expand the capabilities of single devices by allowing them to exchange information and interact with each other.
Molecular communication (MC) has been acknowledged as one of the most promising solutions to the problem of communication in bio-inspired nanonetworks, due to its unique potential benefits of bio\text{-}compatibility and low energy consumption~\cite{Survey}. In MC, the information transmission between devices is realized through the exchange of molecules~\cite{Andrew_Book}.
The simplest molecular propagation mechanism is free diffusion, where the information-carrying molecules can propagate from the transmitter to the receiver via the Brownian motion. Therefore, no external energy is required for diffusion-based propagation.

One of the primary challenges posed by diffusion-based MC is that its reliability quickly decreases when the distance between transceivers increases. In order to boost its reliability, one approach that can be adapted from conventional wireless communications is where
multiple receivers sharing common information are used. It is indeed often in biological environments that small-scale devices or organisms share common information (e.g., odor, flavor, location, and chemical state)
to achieve a specific task \cite{single and multiple}. For example, in the application of drug delivery, one nanoscale device that arrives at a target site (e.g., tumor cells) broadcasts the location of that target site. Other nanoscale devices are then recruited to the target site, thereby enhancing the targeting efficacy \cite{drug delivery}. Another example where organisms share common information is the cooperation in a functional unit of an actin filament. The functional unit consists of one troponin, one tropomyosin, and seven actin monomers. When the troponin responds to changes in the calcium concentration of the medium, the response is propagated to all seven actin monomers. The actin monomers then form ``rigor complexes'' with tropomyosin and troponin binds calcium with greater affinity \cite{filament}.


The majority of existing MC studies in the literature has focused on the modeling of a single MC link. Built upon these studies, some papers such as \cite{Multiple-access broadcast,Silico Experiment,Bacteria,Tranmissiom Rate,Molecular MIMO} have investigated multiple-receiver MC systems. However, the active cooperation between multiple receivers to determine the transmitter's signal has not been considered in the literature.
In \cite{Multiple-access broadcast}, the model of a molecular broadcast channel where a single transmitter transmits molecular information to multiple receivers was developed and the capacity of this channel was analyzed.
In \cite{Silico Experiment}, simulations were performed to demonstrate the feasibility of a bacterium-based bionanosensor network where bacterium-based bionanomachines collectively perform target detection and tracking.
The authors of \cite{Bacteria} studied the communication process between two populations of bacteria through a diffusion channel. Considering the communication between a group of transmitters and a group of receivers, \cite{Tranmissiom Rate} optimized the transmission rate at each transmitter. Very recently, the authors of \cite{Molecular MIMO} designed a multiple-input multiple-output MC system and characterized the inter-symbol and inter-link interference in such a system.

In other fields of communications, e.g., wireless communications, cooperation among multiple distributed detectors has been extensively studied to reveal its benefits in detection performance improvement. For example, in cooperative spectrum sensing, multiple secondary users share sensing data to improve the detection quality of a primary user \cite{cooperative spectrum sensing}. Generally, in a distributed detection system, the data of the individual detectors is shared with a fusion center where the received information is appropriately combined to yield a global inference \cite{Distributed Detection}. This data may be hard decisions, soft decisions (multi-level decisions instead of binary decisions), or quantized observations. We note that MC is a suitable domain to apply distributed detection to improve reliability, but this has not yet been studied.

In this paper, we consider a cooperative diffusion-based MC system in which multiple receivers collaboratively detect a transmitter's bit sequence. This is the first attempt to apply cooperation via distributed detection in the MC domain. Our goal is to demonstrate the increase in reliability by distributing resources over multiple receivers. In our considered system, individual receivers make local hard decisions about each transmitted bit and then report these decisions to a fusion center. The fusion center fuses all local hard decisions to make a global decision using an $N$-out-of-$K$ fusion rule. We also consider asymmetric and symmetric topologies for the system. 
For each topology, we consider two different reporting scenarios, namely, perfect reporting and noisy reporting.
For both reporting scenarios, we derive closed-form analytical expressions for the expected global error probabilities. Using numerical and simulation results, we demonstrate that our analytical expressions are accurate and the error performance of our considered system is significantly better than the point-to-point link which consists of one transmitter and one receiver.


\section{System Model}\label{sec:system model}


\begin{figure}[!t]
\centering
\includegraphics[width=0.9\columnwidth]{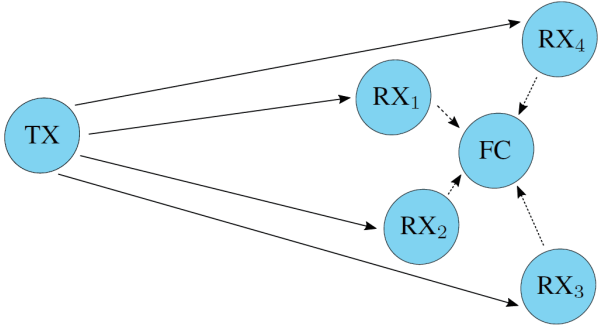}
\caption{Illustration of a cooperative MC system with $K=4$, where the transmission from the TX to the RXs and the decision reporting from the RXs to the FC are represented by solid and dashed arrows, respectively.}
\label{system model}
\end{figure}

We consider a cooperative MC system in a three-dimensional space, as illustrated in Fig.~\ref{system model}, which consists of one transmitter (TX), a ``cluster'' of $K$ receivers (RXs), and one device acting as a fusion center (FC). We clarify that the FC is not included in the set of RXs. We assume that the TX, RXs, and FC are in the nanoscale 
to microscale 
dimensions. 
We also assume that all RXs and the FC are spherical observers. The volume of the $k$th RX, $\RX_k$, where $k\in\{1,2,\ldots,K\}$, and the FC are denoted by $V_{\ss\R_k}$ and $V_{\ss\FC}$, respectively, and radius of the $\RX_k$ and the FC are denoted by $r_{\ss\R_k}$ and $r_{\ss\FC}$, respectively. We further assume that the RXs and the FC are independent passive observers such that molecules can diffuse through them without reacting.

The TX, RXs, and FC in our considered system communicate in three phases. In the first phase, the TX transmits information via type $A_0$ molecules to the RXs through the diffusive channel. The number of released type $A_0$ molecules is denoted by $S_0$. We assume that the movement of different types of molecules is independent, and the movement of individual molecules of the same type is also independent. The type $A_0$ molecules transmitted by
the TX are detected by all RXs. In this work we consider that the TX uses ON/OFF key modulation \cite{M.J.Moore} to convey information, i.e., the TX releases $S_{0}$ molecules of type $A_0$ to convey information bit ``1'', and releases no molecules to convey information bit ``0''. To enable the ON/OFF key modulation, the information transmitted by the TX is encoded into a binary sequence of length $L$, denoted by $\textbf{W}_{\ss\T}=\{W_{\ss\T}[1],W_{\ss\T}[2],\ldots,W_{\ss\T}[L]\}$, where $W_{\ss\T}[j]$, $j\in\{1,\ldots,L\}$, is the $j$th bit transmitted by the TX. We assume that the probability of transmitting ``1'' in the $j$th bit is $P_{1}$ and the probability of transmitting ``0'' in the $j$th bit is $1-P_{1}$, such that $\textrm{Pr}(W_{\ss\T}[j]=1)=P_1$ and $\textrm{Pr}(W_{\ss\T}[j]=0)=1-P_{1}$, where $\textrm{Pr}(\cdot)$ denotes probability.  

In the second phase, each RX makes a local hard decision on each transmitted bit. We denote $\hat{W}_{{\ss\R}_k}[j]$ as the local hard decision on the $j$th transmitted bit at $\RX_k$.
Then, the RXs simultaneously report their local hard decisions to the FC. We assume that $\RX_k$ transmits type $A_{k}$ molecules, which can be detected by the FC.\footnote{We acknowledge that using a unique type of molecule at each RX may not necessarily be a realistic assumption. The rational behind the adoption of this assumption is to give a lower bound on the error performance of the cooperative MC system.} The number of released type $A_k$ molecules is denoted by $S_k$. We also assume that the channel between each RX and the FC is diffusion-based, and each RX uses ON/OFF key modulation \cite{M.J.Moore} to report its local hard decision. For example, if the local hard decision at $\RX_k$ is bit ``1'', $\RX_k$ releases $S_{{k}}$ molecules of type $A_k$ to report it to the FC; otherwise, $\RX_k$ releases no molecules.



In the final phase, the FC obtains the decision at $\RX_k$ by receiving type $A_k$ molecules over the $\RX_{k}-\FC$ link. It is assumed that the $K$ $\RX_{k}-\FC$ links are independent. We denote $\hat{W}_{{\ss\FC}_{k}}[j]$ as the received local decision of $\RX_k$ on the $j$th transmitted bit at the FC. The FC combines all $\hat{W}_{{\ss\FC}_{k}}[j]$ using an $N$-out-of-$K$ fusion rule to make a global decision $\hat{W}_{\ss\FC}[j]$ on the $j$th bit transmitted by the TX. According to the $N$-out-of-$K$ fusion rule, the FC declares a global decision of ``1'' when it receives at least $N$ decisions of ``1''. There are several special cases of the $N$-out-of-$K$ fusion rule: 1) The majority decision rule where $N=\ceil{K/2}$ and $\ceil{x}$ represents the smallest integer greater than or equal to $x$, 2) The OR rule where $N=1$, and 3) The AND rule where $N=K$.
We clarify that we do not consider the reporting of the FC's global decision back to the RXs.
To simplify the notation, we define $\textbf{W}_{{\ss\T}}^{l}=\{W_{\ss\T}[1],\ldots,W_{\ss\T}[l]\}$ as a subsequence of length $l$ transmitted by the TX, where $l\leq{L}$, $\hat{\textbf{W}}_{\RX_k}^l=\{\hat{W}_{\RX_k}[1],\ldots,\hat{W}_{\RX_k}[l]\}$ as a subsequence of the local hard decisions at $\RX_k$, $\hat{\textbf{W}}_{{\FC_k}}^l=\{\hat{W}_{\FC_k}[1],\ldots,\hat{W}_{\FC_k}[l]\}$ as a subsequence of the received local decision of $\RX_k$ at the FC, 
and $\hat{\textbf{W}}_{{\FC}}^l=\{\hat{W}_{\FC}[1],\ldots,\hat{W}_{\FC}[l]\}$ as a subsequence of the global decisions at the FC.


We denote $t_{\trans}$ as the transmission interval time from the TX to the RXs and $t_{\report}$ as the report interval time from the RXs to the FC. As such, the bit interval time from the TX to the FC is given by $T = t_{\trans}+t_{\report}$. At the beginning of the $j$th bit interval, $(j-1)T$, the TX transmits $W_{\ss\T}[j]$. After this the TX keeps silent until the end of the $j$th bit interval.
We assume that the weighted sum detector \cite{Adam optimal} is adopted at the RXs and FC for detection. Thus, the RXs and FC each take multiple samples equally spaced in their corresponding interval time, add the individual samples with a certain weight for each sample, and compare the summation with a decision threshold. The detection threshold at $\RX_k$ and FC are denoted by $\xi_{\ss\R_k}$ and $\xi_{\ss\FC}$, respectively. We assume equal weights for all samples to decrease the computational complexity of the detector and facilitate its usage in MC.

We now detail the sampling schedules of the RXs and FC. Each RX takes $M_{\RX}$ samples in each bit interval at the same time. The time of the $m$th sample for each RX in the $j$th bit interval is given by $t_{\ss\R}(j,m) = (j-1)T + m\Delta{t_{\ss\R}}$, where $\Delta{t_{\ss\R}}$ is the time step between two successive samples at each RX, $m\in\left\{1,2,\ldots,M_{\RX}\right\}$, and $M_{\RX}\Delta{t_{\ss\R}}<t_\trans$. We consider that the RXs operate in half-duplex mode, where they do not receive the information and report their local decisions at the same time. Specifically, at the time $(j-1)T + t_{\trans}$, each RX reports its local decision for the $j$th interval via diffusion to the FC. We assume that the FC is able to simultaneously and independently detect the different types of molecules from all RXs (as in \cite{single and multiple}) and it takes $M_{\FC}$ samples of each type of molecule in every reporting interval. The time of the $\tilde{m}$th sample of type $A_k$ molecules at the FC in the $j$th bit interval is given by $t_{\ss\FC}(j,\tilde{m})=(j-1)T+t_{\trans}+\tilde{m}\Delta{t_{\ss\FC}}$, where $\Delta{t_{\ss\FC}}$ is the time step between two successive samples at the FC and $\tilde{m}\in\left\{1,2,\ldots,M_{\FC}\right\}$.

\section{Error Performance Analysis of Cooperative MC Systems}\label{sec:Performance Analysis}

In this section, we first establish some fundamental preliminary results to facilitate the error performance analysis of the cooperative MC system. Using the preliminary results, we then analyze the expected global error probability of the cooperative MC system.

\subsection{Fundamental Preliminaries}\label{sec:Preliminaries}

In this subsection we examine the expected error probabilities of the $\TX-\RX_k$ link and the $\TX-\RX_k-\textrm{FC}$ link. This examination is based on the analytical methods presented in \cite{multi hop}.

\subsubsection{$\TX-\RX_k$ Link}

We first focus on the $\TX-\RX_k$ link. Given independent molecular behavior and the fact that the RXs are sufficiently far from the TX,
we use \cite[Eq.~(20)]{Adam dimen} to evaluate the probability of observing a given type $A_0$ molecule, emitted from the TX at $t=0$, inside $V_{\ss\R_k}$ at time $t$. Such a probability is given by
\begin{align}\label{probability}
P_{\ob,0}^{({\ss{\T},{\R_k}})}(t) = \frac{V_{\ss\R_k}}{(4\pi D_{0}t)^{3/2}}\exp\left(-\frac{d_{\ss\T_k}^{2}}{4D_{0}t}\right),
\end{align}
where $D_{0}$ is the diffusion coefficient of type $A_0$ molecules in $\frac{\m^{2}}{\s}$ and $d_{\ss\T_k}$ is the distance between TX and $\RX_k$ in $\m$.

We denote $S_{\ob,0}^{({\ss{\T},{\R_k}})}[j]$ as the number of molecules observed within $V_{\ss\R_k}$ in the $j$th bit interval due to the emission of molecules from the current and previous bit intervals at the TX, $\textbf{W}_{\ss\T}^j$. As discussed in \cite{multi hop}, $S_{\ob,0}^{({\ss{\T},{\R_k}})}[j]$ can be accurately approximated by a Poisson random variable with the mean given by
\begin{align}\label{observed molecular numbers R}
\bar{S}_{\ob,0}^{({\ss{\T},{\R_k}})}[j]
= &\;S_{0}\sum\limits^{j}_{i=1}W_{\ss\T}[i]\nonumber\\
&\times\sum\limits^{M_{\ss\RX}}_{m=1}P_{\ob,0}^{({\ss{\T},{\R_k}})}((j-i)T + m\Delta{t_{\ss\R}}).
\end{align}
Then, the decision at $\RX_k$ in the $j$th bit interval is given by
\begin{align}\label{detectorRX}
\hat{W}_{\ss\R_k}[j]=
\begin{cases}
1,&\mbox{if $S_{\ob,0}^{({\ss{\T},{\R_k}})}[j]\geq\xi_{\ss\R_k}$,}\\
0,&\mbox{otherwise}.
\end{cases}
\end{align}
Moreover, based on \cite[Eq.~(9)]{multi hop}, the expected miss detection probability for given $\textbf{W}_{\ss\T}^{j-1}$ in the $j$th bit interval\footnote{For the sake of simplicity, we define $P_{\md,k}[j]\triangleq{}P_{\md,k}[j|\textbf{W}_{\ss\T}^{j-1}]$ in \eqref{Pm1} and $P_{\fa,k}[j]\triangleq{}P_{\fa,k}[j|\textbf{W}_{\ss\T}^{j-1}]$ in \eqref{Pf1}. Similarly, we define $\tilde{P}_{\md,k}[j]\triangleq{}\tilde{P}_{\md,k}[j|\textbf{W}_{\ss\T}^{j-1}]$ in \eqref{Pem2}, $\tilde{P}_{\fa,k}[j]\triangleq{}\tilde{P}_{\fa,k}[j|\textbf{W}_{\ss\T}^{j-1}]$ in \eqref{Pef2}, and $Q_{\ss\FC}[j]\triangleq{}Q_{\ss\FC}[j|\textbf{W}_{\ss\T}^{j-1}]$, $Q_{\md}[j]\triangleq{}Q_{\md}[j|\textbf{W}_{\ss\T}^{j-1}]$, and $Q_{\fa}[j]\triangleq{}Q_{\fa}[j|\textbf{W}_{\ss\T}^{j-1}]$ in all equations in Section \ref{sec:Performance of the Cooperative MC System}.}
of the $\TX-\RX_k$ link is written as
\begin{align}\label{Pm1}
P_{\md,k}[j] = \textrm{Pr}(S_{\ob,0}^{({\ss{\T},{\R_k}})}[j]<\xi_{\ss\R_k}|W_{\ss\T}[j]=1,\textbf{W}_{\ss\T}^{j-1}),
\end{align}
and the corresponding expected false alarm probability is written as
\begin{align}\label{Pf1}
P_{\fa,k}[j] = \textrm{Pr}(S_{\ob,0}^{({\ss{\T},{\R_k}})}[j]\geq\xi_{\ss\R_k}|W_{\ss\T}[j]=0,\textbf{W}_{\ss\T}^{j-1}).
\end{align}

\subsubsection{$\TX-\RX_k-\FC$ Link}

Next, we focus on the $\TX-\RX_k-\FC$ link. We clarify that, due to the FC's intended proximity to $\RX_k$, we cannot use \eqref{probability} to evaluate the probability of observing a given $A_k$ molecule, emitted from the $\RX_k$ at $t=0$, inside $V_{\ss\FC}$ at time $t$, which is denoted by $P_{\ob,{k}}^{({\ss\R_k,\FC})}(t)$. Instead, we apply \cite[Eq.~(27)]{Adam dimen} to derive
$P_{\ob,{k}}^{(\ss\R_k,\ss\FC)}(t)$ 
as
\begin{align}\label{general prob}
P_{\ob,{k}}^{({\ss\R_k,\FC})}(t) = &\;\frac{1}{2}\left[\erf\left(\frac{r_{\ss\FC}+d_{\ss\FC_k}}{2\sqrt{D_{{k}}t}}\right)+\erf\left(\frac{r_{\ss\FC}-d_{\ss\FC_k}}{2\sqrt{D_{{k}}t}}\right)\right]\nonumber\\
&-\frac{\sqrt{D_{{k}}t}}{d_{\ss\FC_k}\sqrt{\pi}}\left[\exp\left(-\frac{(-d_{\ss\FC_k}+r_{\ss\FC})^{2}}{4D_{{k}}t}\right)\right.\nonumber\\
&\left.-\exp\left(-\frac{(-d_{\ss\FC_k}-r_{\ss\FC})^{2}}{4D_{{k}}t}\right)\right],
\end{align}
where $D_{{k}}$ is the diffusion coefficient of type $A_{k}$ molecules in $\frac{\m^{2}}{\s}$ and $d_{\ss\FC_k}$ is the distance between $\RX_k$ and $\FC$ in $\m$.

We denote ${S}_{\ob,{k}}^{({\ss\R_k,\FC})}[j]$ as the number of molecules observed within $V_{\ss\FC}$ in the $j$th bit interval due to the emissions of molecules from the current and the previous bit intervals at $\RX_k$, $\hat{\textbf{W}}_{\RX_k}^j$. We note that the TX and $\RX_k$ use the same key modulation method and the $\TX-\RX_k$ and $\RX_k-\FC$ links are both diffusion-based. Therefore, ${S}_{\ob,{k}}^{({\ss\R_k,\FC})}[j]$ can also be accurately approximated by a Poisson random variable. We denote ${\bar{S}}_{\ob,{k}}^{({\ss\R_k,\FC})}[j]$ as the mean of ${S}_{\ob,{k}}^{({\ss\R_k,\FC})}[j]$ and obtain it by replacing $S_{0}$, $W_{\ss\T}[i]$, $P_{\ob,0}^{({\ss{\T},{\R_k}})}$, $M_{\ss\RX}$, $m$, and $\Delta{t_{\ss\R}}$ with $S_{{k}}$, $\hat{W}_{\ss\R_k}[i]$, $P_{\ob,{k}}^{({\ss\R_k,\FC})}$, $M_{\ss\FC}$, $\tilde{m}$, and $\Delta{t_{\ss\FC}}$ in \eqref{observed molecular numbers R}, respectively.
Similarly, $\hat{W}_{{\ss\FC}_{k}}[j]$ can be obtained by replacing $S_{\ob,0}^{({\ss{\T},{\R_k}})}[j]$ and $\xi_{\ss\R_k}$ with ${S}_{\ob,{k}}^{({\ss\R_k,\FC})}[j]$ and $\xi_{\ss\FC}$ in \eqref{detectorRX}, respectively. Furthermore, based on \cite[Eqs.~(13) and (14)]{multi hop}, the expected miss detection probability for given $\textbf{W}_{\ss\T}^{j-1}$ in the $j$th bit interval of the $\TX-\RX_k-\textrm{FC}$ link is derived as
\begin{align}\label{Pem2}
\tilde{P}_{\md,k}[j]=&\;\textrm{Pr}(S_{\ob,{0}}^{({\ss{\T},{\R_k}})}[j]\geq\xi_{\ss\R_k}|W_{\ss\T}[j]=1,\textbf{W}_{\ss\T}^{j-1})\nonumber\\
                             &\times\textrm{Pr}(S_{\ob,{k}}^{({\ss\R_k,\FC})}[j]<\xi_{\ss\FC}|\hat{W}_{\ss\R_k}[j]=1,\hat{\textbf{W}}_{\ss\R_k}^{j-1})\nonumber\\
                             &+\textrm{Pr}(S_{\ob,{0}}^{({\ss{\T},{\R_k}})}[j]<\xi_{\ss\R_k}|W_{\ss\T}[j]=1,\textbf{W}_{\ss\T}^{j-1})\nonumber\\
                             &\times\textrm{Pr}(S_{\ob,{k}}^{({\ss\R_k,\FC})}[j]<\xi_{\ss\FC}|\hat{W}_{\ss\R_k}[j]=0,\hat{\textbf{W}}_{\ss\R_k}^{j-1}),
\end{align}
and the corresponding expected false alarm probability is derived as
\begin{align}\label{Pef2}
\tilde{P}_{\fa,k}[j]=&\;\textrm{Pr}(S_{\ob,{0}}^{({\ss{\T},{\R_k}})}[j]\geq\xi_{\ss\R_k}|W_{\ss\T}[j]=0,\textbf{W}_{\ss\T}^{j-1})\nonumber\\
                             &\times\textrm{Pr}(S_{\ob,{k}}^{({\ss\R_k,\FC})}[j]\geq\xi_{\ss\FC}|\hat{W}_{\ss\R_k}[j]=1,\hat{\textbf{W}}_{\ss\R_k}^{j-1})\nonumber\\
                             &+\textrm{Pr}(S_{\ob,{0}}^{({\ss{\T},{\R_k}})}[j]<\xi_{\ss\R_k}|W_{\ss\T}[j]=0,\textbf{W}_{\ss\T}^{j-1})\nonumber\\
                             &\times\textrm{Pr}(S_{\ob,{k}}^{({\ss\R_k,\FC})}[j]\geq\xi_{\ss\FC}|\hat{W}_{\ss\R_k}[j]=0,\hat{\textbf{W}}_{\ss\R_k}^{j-1}).
\end{align}

\subsection{Error Performance Analysis}\label{sec:Performance of the Cooperative MC System}

In this subsection, we analyze the expected global error probability of the cooperative MC system. We assume that $\textbf{W}_{\ss\T}^{j-1}$ is given and there is no \emph{a priori} knowledge of $W_{\ss\T}[j]$. As such, the expected global error probability in the $j$th bit interval, $Q_{\ss\FC}[j]$, is written as
\begin{align}\label{overall probability}
Q_{\ss\FC}[j] = P_1Q_{\md}[j] + \left(1-P_{1}\right)Q_{\fa}[j],
\end{align}
where $Q_{\md}[j]$ and $Q_{\fa}[j]$ are the expected global miss detection probability in the $j$th bit interval and the expected global false alarm probability in the $j$th bit interval, respectively. The expected average global error probability, $\overline{Q}_{\ss\FC}$, is obtained by averaging $Q_{\ss\FC}[j]$ over all possible realizations of  $\textbf{W}_{\ss\T}^{j-1}$ and across all bit intervals.

In the following, we evaluate $Q_{\md}[j]$ and $Q_{\fa}[j]$ for the asymmetric and symmetric topologies. In the asymmetric topology, the distances between the TX and the RXs are non-identical and/or the distances between the RXs and the FC are non-identical. In the symmetric topology, the distances between the TX and the RXs are identical and the distances between the RXs and the FC are also identical. For each topology, we consider two different reporting scenarios, namely, perfect reporting and noisy reporting. In the perfect reporting scenario, we assume that no error occurs when $\RX_k$ reports to the FC, i.e., $\hat{W}_{{\ss\FC}_{k}}[j]= \hat{W}_{{\ss\R}_k}[j]$. In the noisy reporting scenario, we take into consideration the errors in the reporting from $\RX_k$ to the FC due to diffusion. In addition, we clarify that the bit interval time, the number of molecules for bit ``1'' released by the TX, and the sampling schedules of the RXs and the FC are the same in the perfect and noisy reporting scenarios.

\subsubsection{Asymmetric Topology}\label{sec:Asymmetrical}

In the asymmetric topology, the RXs have independent and \emph{non-identically} distributed observations. Thus, in order to evaluate $Q_{\md}[j]$ and $Q_{\fa}[j]$ for given $\textbf{W}_{\ss\T}^{j-1}$, we need to evaluate the expected miss detection probabilities and the expected false alarm probabilities of the $\TX-\RX_k$ and $\TX-\RX_k-\textrm{FC}$ links for each RX, i.e., $P_{\md,k}[j]$, $P_{\fa,k}[j]$, $\tilde{P}_{\md,k}[j]$, and $\tilde{P}_{\fa,k}[j]$.

We first consider the perfect reporting scenario. Since no error occurs when $\RX_k$ reports to the FC, we use $P_{\md,k}[j]$ and $P_{\fa,k}[j]$, given by \eqref{Pm1} and \eqref{Pf1}, respectively, to evaluate $Q_{\md}[j]$ and $Q_{\fa}[j]$. To facilitate this evaluation for the $N$-out-of-$K$ fusion rule, we first define a set $\mathcal{R}$ which includes $K$ RXs. As such, there are $\binom{K}{n}$ subsets of $n$ RXs that can be taken from $K$ RXs, where $N\leq{n}\leq{K}$. We then denote $\mathcal{A}_q$ as one such subset and $\mathcal{R}\setminus\mathcal{A}_q$ as the set containing the remaining $K-n$ RXs, where $q\in\left\{1,2,\ldots,\binom{K}{n}\right\}$.
Therefore, we derive $Q_{\md}[j]$ and $Q_{\fa}[j]$ as
\begin{align}\label{aQm_KN}
Q_{\md}[j] = 1-\sum\limits^{K}_{n=N}\sum\limits^{\binom{K}{n}}_{q=1}\prod_{k\in{\mathcal{A}_q}}\left(1-P_{\md,k}[j]\right)
\prod_{k\in\{\mathcal{R}/\mathcal{A}_q\}}P_{\md,k}[j]
\end{align}
and
\begin{align}\label{aQf_KN}
Q_{\fa}[j] = \sum\limits^{K}_{n=N}\sum\limits^{\binom{K}{n}}_{q=1}\prod_{k\in{\mathcal{A}_q}}P_{\fa,k}[j]
\prod_{k\in\{\mathcal{R}/\mathcal{A}_q\}}\left(1-P_{\fa,k}[j]\right),
\end{align}
respectively. For the OR rule, 
we obtain $Q_{\md}[j]$ and $Q_{\fa}[j]$ as
\begin{align}\label{aQm_OR}
Q_{\md}[j] = \prod\limits^{K}_{k=1}P_{\md,k}[j]
\end{align}
and
\begin{align}\label{aQf_OR}
Q_{\fa}[j] = 1-\prod\limits^{K}_{k=1}\left(1-P_{\fa,k}[j]\right),
\end{align}
respectively. For the AND rule, 
we obtain $Q_{\md}[j]$ and $Q_{\fa}[j]$ as
\begin{align}\label{aQm_AND}
Q_{\md}[j] = 1-\prod\limits^{K}_{k=1}\left(1-P_{{\md},k}[j]\right)
\end{align}
and
\begin{align}\label{aQf_AND}
Q_{\fa}[j] = \prod\limits^{K}_{k=1}P_{{\fa},k}[j],
\end{align}
respectively. We highlight that the expressions derived in \eqref{aQm_KN}--\eqref{aQf_AND} are in closed form.

In the noisy reporting scenario, we consider errors in the reporting from $\RX_k$ to the FC due to diffusion. As such, we use $\tilde{P}_{{\md},k}[j]$ and $\tilde{P}_{{\fa},k}[j]$, given by \eqref{Pem2} and \eqref{Pef2}, respectively, to evaluate $Q_{\md}[j]$ and $Q_{\fa}[j]$. Specifically, $Q_{\md}[j]$ and $Q_{\fa}[j]$ for the $N$-out-of-$K$ rule, OR rule, and AND rule in the noisy reporting scenario are obtained by replacing $P_{{\md},k}[j]$ and $P_{{\fa},k}[j]$ with $\tilde{P}_{{\md},k}[j]$ and $\tilde{P}_{{\fa},k}[j]$, respectively, in \eqref{aQm_KN}--\eqref{aQf_AND}.


\subsubsection{Symmetric Topology}\label{sec:Symmetrical}

In the symmetric topology, RXs have independent and \emph{identically} distributed observations. Thus, for the $\T-\R_k$ link, we write $P_{{\md},k}[j]= P_{\md}[j]$ and $P_{{\fa},k}[j]= P_{\fa}[j]$. For the $\T-\R_k-\FC$ link, we write $\tilde{P}_{{\md},k}[j]= \tilde{P}_{\md}[j]$ and $\tilde{P}_{{\fa},k}[j]= \tilde{P}_{\fa}[j]$.

Again, let us first consider the perfect reporting scenario. For the $N$-out-of-$K$ fusion rule, 
$Q_{\md}[j]$ and $Q_{\fa}[j]$ are simplified as
\begin{align}\label{Qm_KN}
Q_{\md}[j]= 1-\sum\limits^{K}_{n=N}\binom{K}{n}\left(1-P_{{\md}}[j]\right)^{n}{P_{{\md}}[j]}^{K-n}
\end{align}
and
\begin{align}\label{Qf_KN}
Q_{\fa}[j] = \sum\limits^{K}_{n=N}\binom{K}{n}{P_{f}[j]}^{n}\left(1-P_{{\fa}}[j]\right)^{K-n},
\end{align}
respectively. For the OR rule, 
$Q_{\md}[j]$ and $Q_f[j]$ are simplified as
\begin{align}\label{Qm_OR}
Q_{\md}[j] = P_{{\md}}[j]^K, \quad Q_{\fa}[j] = 1-(1-P_{\fa}[j])^K,
\end{align}
respectively. For the AND rule, 
$Q_{\md}[j]$ and $Q_{\fa}[j]$ are simplified as
\begin{align}\label{Qm_AND}
Q_{\md}[j] = 1-(1-P_{{\md}}[j])^K, \quad Q_{\fa}[j] = P_{{\fa}}[j]^K,
\end{align}
respectively. Furthermore, we focus on the noisy reporting scenario for the symmetric topology. In this scenario, we obtain $Q_{\md}[j]$ and $Q_{\fa}[j]$ for the $N$-out-of-$K$ rule, OR rule, and AND rule by replacing $P_{\md}[j]$ and $P_{\fa}[j]$ with $\tilde{P}_{\md}[j]$ and $\tilde{P}_{\fa}[j]$, respectively, in \eqref{Qm_KN}--\eqref{Qm_AND}.



\section{Numerical Results and Simulations}\label{sec:Numerical}


In this section, we present numerical and simulation results to examine the error performance of the considered cooperative MC system. In this examination we use a particle-based stochastic simulator. We list all the environmental parameters adopted in the examination in Table~\ref{tab:table1} and keep them fixed throughout this section. The only parameters that we vary are the detection threshold at $\RX_k$, $\xi_{\ss\R_k}$, the detection threshold at the FC, $\xi_{\ss\FC}$, and the number of RXs, $K$.

\begin{table}[!t]
\centering
\caption{Environmental Parameters Used in Section~\ref{sec:Numerical}}\label{tab:table1}
\begin{tabular}{|c|c|c|}
\hline
Parameter & Symbol& Value \\\hline
Radius of RXs& $r_{\ss\R_k}$ & $0.225\,{\mu}\metre$\\\hline
Radius of FC & $r_{\ss\FC}$ & $0.225\,{\mu}\metre$ \\\hline
Time step at RX & $\Delta{t_{\ss\R}}$ & $100\,{\mu}\s$\\\hline
Time step at FC & $\Delta{t_{\ss\FC}}$ & $10\,{\mu}\s$ \\\hline
Number of samples of RX& $M_{\RX}$ & 5 \\\hline
Number of samples of FC& $M_{\FC}$ & 5 \\\hline
Transmission time interval & $t_{\trans}$ & $1\,{\m}\s$\\\hline
Report time interval & $t_{\report}$ & $0.1\,{\m}\s$\\\hline
Bit interval time& $T$ & $1.1\,{\m}\s$\\\hline
Diffusion coefficient & $D_0=D_{k}$ & $5\times10^{-9}{\m^{2}}/{\s}$\\\hline
Length of transmitter sequence & $L$ & $10$ \\\hline
Probability of binary 1 & $P_1$ & $0.5$ \\
\hline
\end{tabular}
\end{table}

\begin{table}[!t]
\centering
\caption{Devices' Location for the Symmetric Topology}\label{tab:coordinates1}
\begin{tabular}{|c|c|c|c|}
\hline
Devices & X-axis [${\mu}\metre$]& Y-axis [${\mu}\metre$] & Z-axis [${\mu}\metre$]\\\hline
$\TX$ & 0 & 0 & 0\\\hline
$\RX_1$ & 2 & 0.6 & 0\\\hline
$\RX_2$ & 2 & -0.6& 0 \\\hline
$\RX_3$ & 2 & 0& 0.6 \\\hline
$\RX_4$ & 2 & 0& -0.6 \\\hline
$\RX_5$ & 2 & 0.3 & 0.5196 \\\hline
$\RX_6$ & 2 & 0.3 & -0.5196 \\\hline
$\FC$   & 2 & 0   & 0 \\
\hline
\end{tabular}

\end{table}

\begin{table}[!t]
\centering
\caption{Devices' Location for the Asymmetric Topology}\label{tab:coordinates 2}
\begin{tabular}{|c|c|c|c|}
\hline
Devices & X-axis [${\mu}\metre$]& Y-axis [${\mu}\metre$] & Z-axis [${\mu}\metre$]\\\hline
$\TX$   & 0 & 0 & 0 \\\hline
$\RX_1$ & 1.5 & 0.6 & 0\\\hline
$\RX_2$ & 2 & 0.6& 0 \\\hline
$\RX_3$ & 2.5 & 0.6 & 0 \\\hline
$\FC$ & 2 & 0 & 0\\
\hline
\end{tabular}
\end{table}

\begin{figure}[!t]
\centering
\includegraphics[height=2.4in]{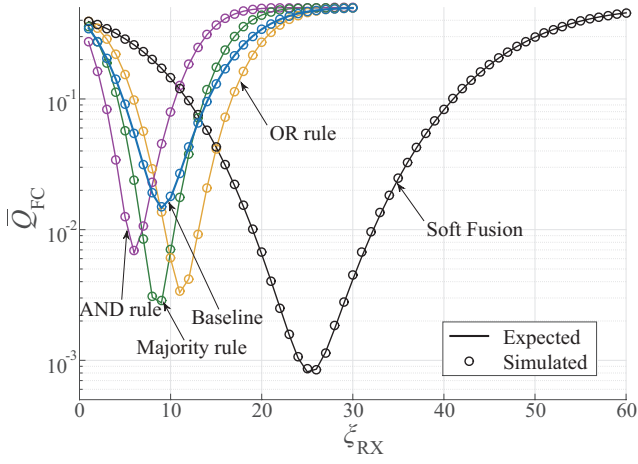}
\caption{Average global error probability $\overline{Q}_{\ss\FC}$ versus the detection threshold at RXs $\xi_{\ss\R}$ with $K=3$ in the perfect reporting scenario.}
\label{Perfect Report}
\end{figure}

\begin{figure}[!t]
\centering
\includegraphics[height=2.4in]{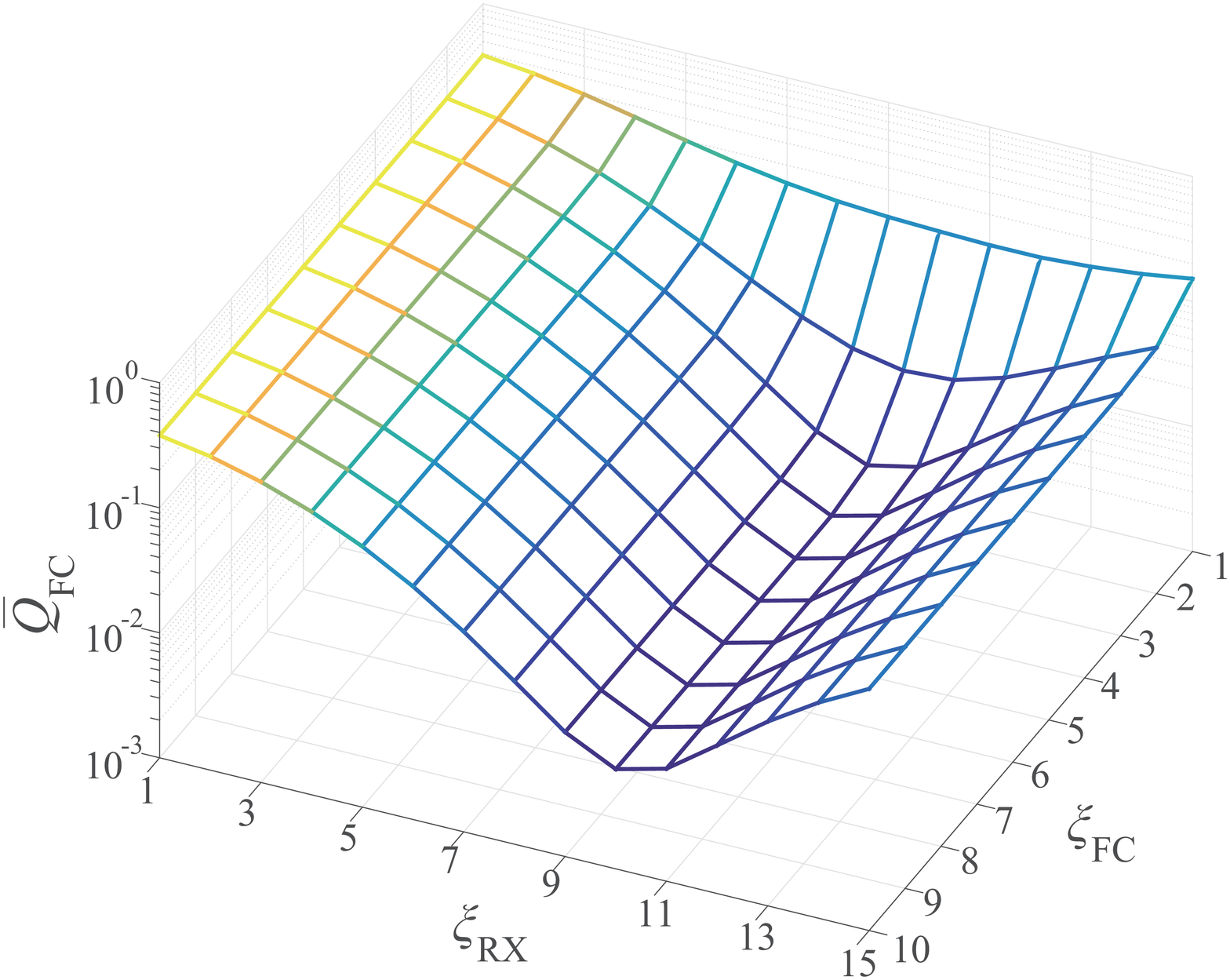}
\caption{Expected average global error probability $\overline{Q}_{\ss\FC}$ versus the detection threshold at RXs $\xi_{\ss\R}$ and the detection threshold at the FC $\xi_{\ss\FC}$ with $K = 2$ in the noisy reporting scenario.}
\label{Pe3D}
\end{figure}

\begin{figure}[!t]
\centering
\includegraphics[height=2.4in]{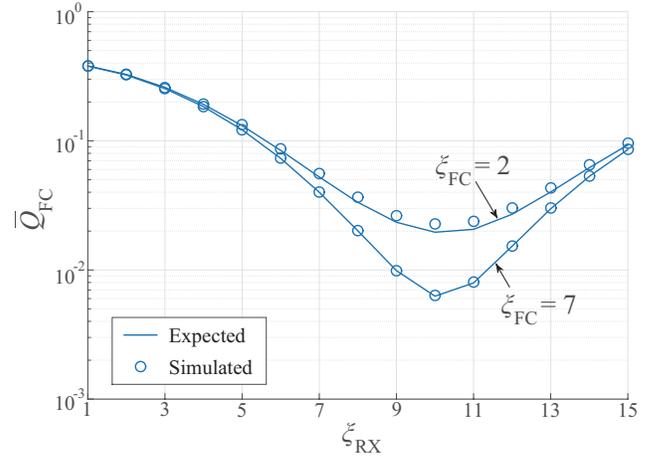}
\caption{Average global error probability $\overline{Q}_{\ss\FC}$ versus the detection threshold at RXs $\xi_{\ss\R}$ with $K = 2$ 
in the noisy reporting scenario.}
\label{simulation3D}
\end{figure}

\begin{figure}[!t]
\centering
\includegraphics[height=2.4in]{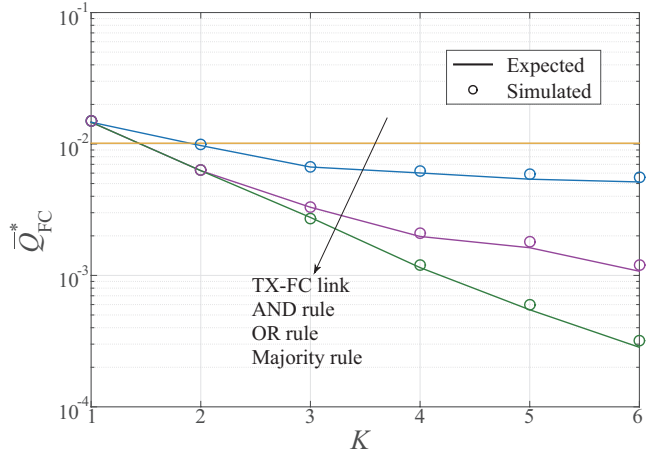}
\caption{Optimal average global error probability $\overline{Q}_{\ss\FC}^{\ast}$ of different fusion rules versus the number of cooperative RXs $K$ in the noisy reporting scenario. The case with $K=1$ is referred to as the baseline case.}
\label{Pe_k}
\end{figure}

\begin{figure}[!t]
\centering
\includegraphics[height=2.4in]{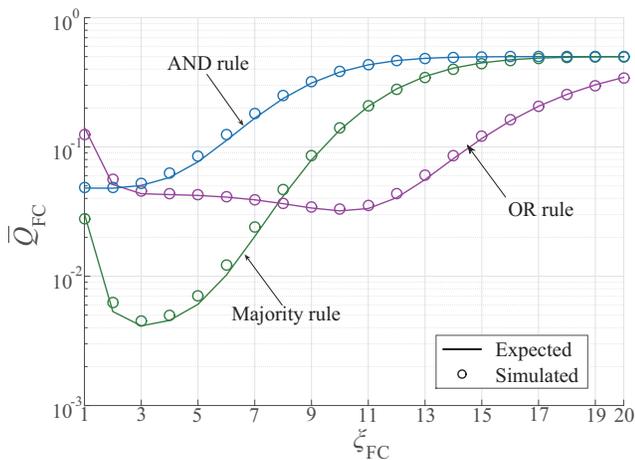}
\caption{Average global error probability $\overline{Q}_{\ss\FC}$ of different fusion rules versus the detection threshold at the FC $\xi_{\ss\FC}$ with the optimal threshold at each RX and $K=3$ in the noisy reporting scenario.}
\label{Pe_thresFC}
\end{figure}

In the following, we assume that the TX releases $S_{0} = 10000$ molecules for information bit ``1'' and the total number of molecules released by all RXs for bit ``1'' is fixed at $1000$, i.e., each RX releases $S_{k}=1000/K$ molecules to report its decision of bit ``1''. Moreover, in Figs.~\ref{Perfect Report}--\ref{Pe_k} we consider a \emph{symmetric} topology that consists of at most six RXs. The specific locations of the TX, RXs, and FC in the symmetric topology are listed in Table~\ref{tab:coordinates1}. Furthermore, in Figs.~\ref{Perfect Report}--\ref{Pe_k} we consider the same detection threshold at the RXs such that $\xi_{\RX_i} = \xi_{\RX}, \forall i$. In Fig.~\ref{Pe_thresFC} we consider an \emph{asymmetric} topology that consists of three RXs. The specific locations of the TX, RXs, and FC in the asymmetric topology are listed in Table~\ref{tab:coordinates 2}.

We compare the error performance of the considered cooperative MC system with that of two point-to-point links, in order to show the performance advantage of the investigated fusion rules. The first point-to-point link is a single $\T-\R$ link, referred to as the baseline case in this section, where only one RX exists but no FC exists. The second point-to-point link is the direct link between the TX and the FC.
In the baseline case, the RX is located at $(2\,{\mu}\metre,0.6\,{\mu}\metre,0)$. In the direct $\TX-\textrm{FC}$ link, the FC is located at $(2\,{\mu}\metre,0,0)$. We assume that in both point-to-point links the TX releases $11000$ molecules, the time step between two successive samples is $100\,{\mu}\s$, and the bit interval time is $T= 1.1\,{\m}\s$. As such, we consider that the total number of molecules, the distance away from the TX, and the bit interval time for the point-to-point links are the same as those for the cooperative MC system listed in Table~\ref{tab:table1}, ensuring the fairness of the comparison.

In Fig.~\ref{Perfect Report}, we consider the \emph{perfect} reporting scenario and plot the average global error probability of a three-RX cooperative system, i.e., $K = 3$, versus the detection threshold at the RXs for the AND rule, OR rule, and majority rule with $N=2$. We see that the three-RX system outperforms the baseline case for all fusion rules. We also see that the majority rule outperforms the OR rule and the OR rule outperforms the AND rule at their corresponding optimal detection thresholds. Furthermore, we consider the performance of a simple soft fusion scheme as a performance bound on our considered system. In this scheme the observation at the FC is equal to the summation of the \emph{observations} at all RXs, rather than their local hard decisions. We find that this scheme has an improved error performance over the AND, OR, and majority rules. This is due to the fact that the three fusion rules are hard fusion schemes where the global decision at the FC is made by integrating local hard decisions made by RXs. We also note that this finding is consistent with the conclusion of distributed detection in other fields of communications, e.g., \cite{soft fusion}.

In Fig.~\ref{Pe3D}, we consider the \emph{noisy} reporting scenario. We plot the expected average global error probability of a two-RX cooperative system, i.e., $K = 2$, versus the detection threshold at the RXs and the detection threshold at the FC for the OR rule. We clearly observe that both $\xi_{\ss\R}$ and $\xi_{\ss\FC}$ affect $\overline{Q}_{\ss\FC}$. Notably, figures such as this one enable us to numerically find the optimal detection thresholds at the RXs and FC that minimize $\overline{Q}_{\ss\FC}$. 
We denote $\overline{Q}_{\ss\FC}^{\ast}$ as the minimum $\overline{Q}_{\ss\FC}$. We clarify from Fig.~\ref{Pe3D} that $\overline{Q}_{\ss\FC}^{\ast}=6.3\times10^{-3}$ when $\xi_{\ss\R} = 10$ and $\xi_{\ss\FC} = 7$. Considering the same parameters, in Fig.~\ref{simulation3D} we plot the average global error probability versus the detection threshold at the RXs for different detection thresholds at the FC. We see that there exists an optimal $\xi_{\RX}$ that minimizes $\overline{Q}_{\ss\FC}$ for a given $\xi_{\FC}$. We further provide the simulation results in Fig.~\ref{simulation3D}, based on which we confirm the accuracy of our expected results.

In Fig.~\ref{Pe_k}, we consider the \emph{noisy} reporting scenario and plot the average global error probability versus the number of cooperative RXs for the AND, OR, and majority rules. This figure highlights the performance advantage of the three fusion rules relative to 1) the baseline case (i.e., TX-RX link) with $K =1$ and 2) the direct TX-FC link. Here, we clarify that in this figure, the \emph{total} number of molecules released by all RXs for information bit ``1'' is fixed for different $K$. We also clarify that in this figure the value of $\overline{Q}_{\ss\FC}^{\ast}$ for each $K$ is the minimum $\overline{Q}_{\ss\FC}$ achieved by numerically optimizing $\xi_{\ss\R}$ and $\xi_{\ss\FC}$, e.g., as in Fig.~\ref{Pe3D} for $K=2$ and the OR rule.
In Fig.~\ref{Pe_k}, we observe that the system error performance profoundly improves as $K$ increases. This is due to the fact that an increasing number of cooperative RXs provide more independent observations of the information bit transmitted by the TX. With the aid of more observations, the probability that all RXs fail to detect the transmitted bit is reduced. We highlight that this observation is consistent with distributed detection in spectrum sensing, e.g., \cite{cooperative spectrum sensing}. We also observe from Fig.~\ref{Pe_k} that, for the same number of RXs, the majority rule outperforms the OR rule and the OR rule outperforms the AND rule. This observation is consistent with that in the perfect reporting scenario in Fig.~\ref{Perfect Report}. We further observe that the cooperative MC system outperforms the TX-RX and TX-FC links for all fusion rules, although the distance of the TX-RX and TX-FC links are shorter than that of the system.
In addition, we observe that the rate of error performance improvement decreases when $K$ increases, especially for the AND rule. This is due to the constraint on the total number of molecules, which means that the number of molecules released by each RX decreases for increasing $K$. It follows that the reporting from the RXs to the FC becomes increasingly unreliable. 


In Fig.~\ref{Pe_thresFC}, we consider a three-RX cooperative system, i.e., $K = 3$, with the \emph{asymmetric} topology in the \emph{noisy} reporting scenario. We plot the average global error probability versus the detection threshold at the FC for the AND, OR, and majority rules. Given that the number of molecules transmitted by the TX is $S_0=10000$, in this figure we consider the optimal threshold at each RX, i.e., $\xi_{{\ss\R}_1}=6$, $\xi_{{\ss\R}_2}=8$, and $\xi_{{\ss\R}_3}=11$. The figure shows that the majority rule outperforms the OR rule and the OR rule outperforms the AND rule at their optimal detection thresholds. This phenomenon is consistent with that in the perfect reporting scenario in Fig.~\ref{Perfect Report} and that in the noisy reporting scenario in Fig.~\ref{Pe_k}. Moreover, we notice that the OR rule outperforms the majority rule and the AND rule when $\xi_{\ss\FC}\geq{8}$, which is consistent with that in the perfect reporting scenario in Fig.~\ref{Perfect Report}. This phenomenon can be explained by the fact that the FC has a lower likelihood to receive an RX's decision of ``1'' when $\xi_{\ss\FC}$ becomes higher. It follows that the OR rule has the best performance among the three fusion rules when $\xi_{\ss\FC}$ is relatively large.


\section{Conclusions}\label{sec:con}

In this paper, we considered a cooperative diffusion-based MC system in which distributed receivers collaboratively determine a transmitter's signal with the aid of a fusion center. For perfect and noisy reporting scenarios, we derived closed-form analytical expressions for the expected global error probability of the system. Our numerical and simulation results showed that the system reliability can be significantly enhanced by combining the detection information of distributed receivers, even when the total number of transmitted molecules is limited. In our future work, we will consider the use of the same type of molecule at each receiver and perform a comprehensive analysis for soft fusion schemes.


\end{document}